\documentclass[twocolumn,showpacs,preprintnumbers,amsmath,amssymb]{revtex4}

\usepackage{graphicx}
\usepackage{dcolumn}
\usepackage{bm}

\begin{document}

\preprint{APS/123-QED}

\title{Superconductivity of the Sr$_2$Ca$_{12}$Cu$_{24}$O$_{41}$ spin ladder system :
Are the superconducting pairing and the spin-gap formation of the
same origin?  }

\author{Naoki Fujiwara }

\altaffiliation{ Email: naokif@issp.u-tokyo.ac.jp}

\author{Nobuo M$\hat{o}$ri}%

\altaffiliation{ Department of Physics, Faculty of Science,
Saitama University, 255 Simookubo, Saitama 338-8581, Japan}%

\author{Yoshiya Uwatoko}
\affiliation {Institute for Solid State Physics, University of
Tokyo, 5-1-5 Kashiwanoha, Kashiwa, Chiba
 277-8581,Japan}

\author{Takehiko Matsumoto}
\affiliation{National Institute for Materials Science, Tsukuba
305-0047, Japan}
\author{Naoki Motoyama and Shinichi Uchida}
\affiliation{Department of Superconductivity, University of
Tokyo,7-3-1 Hongo, Bunkyo-ku, Tokyo 113-8656, Japan }

\date{July 17 2002}

\begin{abstract}
Pressure-induced superconductivity in a spin-ladder cuprate
Sr$_2$Ca$_{12}$Cu$_{24}$O$_{41}$ has not been studied on a
microscopic level so far although the superconductivity was
already discovered in 1996. We have improved high-pressure
technique with using a large high-quality crystal, and succeeded
in studying the superconductivity using $^{63}$Cu nuclear magnetic
resonance (NMR). We found that anomalous metallic state reflecting
the spin-ladder structure is realized and the superconductivity
possesses a s-wavelike character in the meaning that a finite gap
exists in the quasi-particle excitation: At pressure of 3.5GPa we
observed two excitation modes in the normal state from the
relaxation rate $T_1^{-1}$. One gives rise to an activation-type
component in $T_1^{-1}$, and the other $T$-linear component
linking directly with the superconductivity. This gapless mode
likely arises from free motion of holon-spinon bound states
appearing by hole doping, and the pairing of them likely causes
the superconductivity.
\end{abstract}

\maketitle

 Sr$_{14-x}$Ca$_{x}$Cu$_{24}$O$_{41}$ (x=11.5-13.5)
is a spin-ladder system in which superconducting state is realized
by applying pressure [1]. The system possesses a structural unit
of the Cu$_2$O$_3$ two-leg ladder [2], and holes are transferred
from the CuO$_2$ chain unit by substituting Ca for Sr. The dopant
hole density can be controlled over a wide range from 0.07 (x=0)
to 0.24 (x=14) per ladder Cu [3]. It is known from theoretical
investigations that the ground state of undoped system is a
quantum spin liquid and this state persists even in highly doped
region [4-6]. The spin gap has been observed by a number of works
in the present ladder system. The decrease of the spin gap for the
Ca substitution was observed in the NMR measurements [7-9],
although no change was observed in the neutron scattering [10].
The relationship and interplay between the spin gap and the
superconductivity in the hole-doped ladder system is of central
concern [4-6, 11] as is the case with high-$T_{c}$ cuprates in
underdoped region [12].

The superconducting state is realized when high pressure of 3-8GPa
is applied for highly doped compounds (x$\geq$10) [3]. Pressure
plays a role of stabilizing the metallic state and suppressing
anisotropy within the ladder plane. In fact, the temperature ($T$)
dependence of the resistivity in the direction perpendicular to
the plane $\rho_c$ which is insulating at low pressures turns to
be meallic as pressure increases [13, 14]. The ratio of the
resistivities along the rungs and the legs, $\rho$$_a$/$\rho$$_c$
goes up to 80 at ambient pressure, but is reduced below 30 at
3.5GPa [13].

The superconductivity has been studied on a macroscopic level by
the resistivity and AC susceptibility measurements because these
measurements are performable with a small amount of sample and
cubic-anvil pressure cell is available to reach high pressure [1,
13-15]. A clamp-type pressure cell is much more convenient and
available to various methods in which a larger sample volume is
needed. However, usual clamp-type pressure cell is made of CuBe
alloy and the maximum pressure is 3GPa at most. Hence, microscopic
study of the superconductivity has not been performed so far
although the superconductivity was discovered in 1996. Mayaffre
{\it et al.} made NMR measurements under high pressure up to 3GPa
for the field ($H$) perpendicular to the plane, crystal b axis and
suggested that the spin gap is suppressed by applying pressure
[16, 17]. However, the measurements were in fact done not in the
superconducting state but in the normal state above $T_{c}$.

In the present work, we have used a clamp-type pressure cell made
of Ni alloy and performed NMR measurement at pressures more than
3GPa by applying the field parallel to the leg direction, crystal
c axis. This enabled us to study the superconducting state on a
microscopic level for the first time and to investigate pairing
symmetry as well as the relation between the spin gap and the
superconductivity.

Single crystal of Sr$_2$Ca$_{12}$Cu$_{24}$O$_{41}$ with a volume
of 4x2x1mm was prepared for the measurements. The clamp-type
pressure cell with an effective sample space of $\phi$4 x 20mm was
used for the measurement. The appearance of the superconductivity
was confirmed by measuring resonance frequency of a NMR probe
attached to the pressure cell. The resonance frequency is roughly
given as $f\propto 1/ \sqrt{LC}$ where $L$ and $C$ represent
inductance and variable capacitance of the NMR probe,
respectively. The sample is contained in a coil and the onset of
superconductivity is detected by the change of the resonant
frequency, {\it i.e.} the change of $L$ value. This method
corresponds to AC susceptibility measurement. The $T$ dependence
of the frequency at several fields is shown in Fig. 1. If we probe
the temperature at which the resonance frequency starts to change
against $H$, this gives $H_{c2}$ vs. $T_{c}$ characteristics as is
shown in the inset. For high-$T_{c}$ cuprates, the temperature has
been identified as irreversibility temperature $T_{irr}$ [18].
$T_{irr}$ gives a borderline between creeping and freezing of flux
lines (FL) which are formed through the planes for the $H$
perpendicular to the plane. In the present case flux lines are
self-trapped between the planes since the field is applied
parallel to the plane, and thus such a creep is hardly expected.

\begin{figure}
\includegraphics{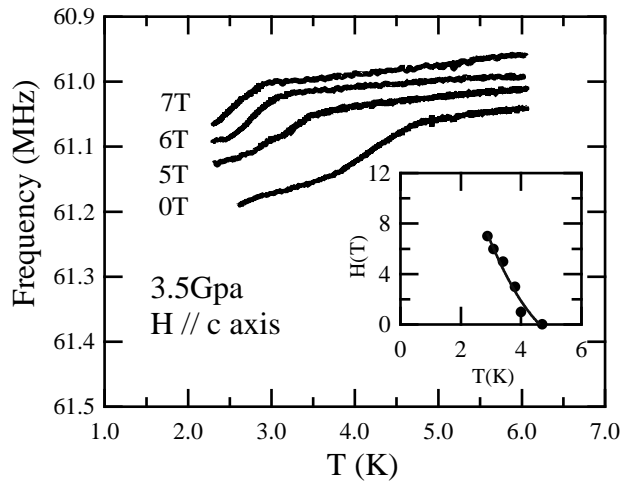}
\caption{\label{fig:epsart}  Resonance frequency of a NMR probe
attached to a pressure cell. The frequency changes at the
superconducting state. The onset corresponds to the upper critical
field. Inset shows $H_{c}-T_{c}$ curve obtained from the onset of
the frequency at each field. Solid line is a guide for the eyes.}
\end{figure}
 $^{63}$Cu-NMR signals for the ladder site can be separated
from those for the chain site since nuclei of these sites possess
different quadrupole coupling [8, 19]. We have reported NMR
spectra under pressure measured by using the present crystal in
Ref. 20. The NMR spectra at 3.5GPa were almost the same with those
at ambient pressure. One signal corresponding to the central
transition (I= -1/2$\Leftrightarrow$ 1/2) of $^{63}$Cu nuclei was
observed for the ladder site and the relaxation rate ($T_1^{-1}$)
was measured at 70.0MHz which corresponds to 6.2T. $T_{c}$ at this
field is about 2.8K as is seen from the inset of Fig. 1. The $T$
dependence of $T_1^{-1}$ is shown in Fig. 2. The spin gap is
observed even under high pressure as an activated $T$ dependence
of $T_1^{-1}$ at temperatures higher than 30K, {\it i.e.},
\begin{equation}
 T_1^{-1}  \propto exp ( -\Delta _{spin} /T).
\end{equation}
The value of $\Delta_{spin}$ is estimated to be 173K.  It should
be noted that the spin gap is seen in the state in which the
charge transport is metallic [13].
\begin{figure}
\includegraphics{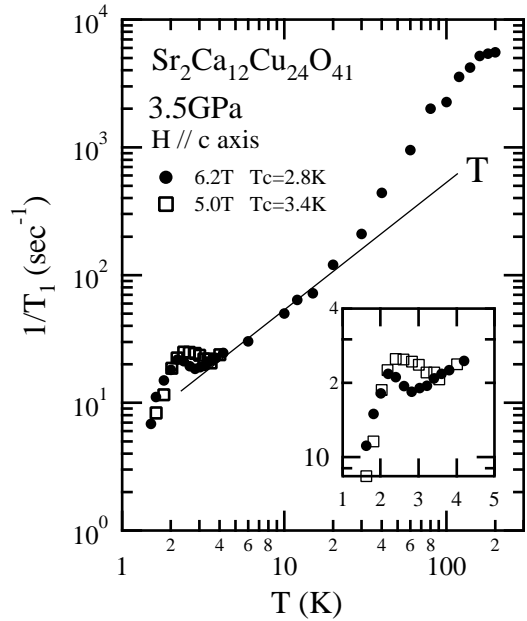}
\caption{\label{fig:epsart} Nuclear ratice relaxation rate $1/T_1$
for $^{63}$Cu nuclei. A solid line shows Korringa relation
expressed by Eq. (2). $1/T_1$ at low temperatures around $T_{c}$
is expanded in the inset.}
\end{figure}

 By contrast,
$T_1^{-1}$ below 30K is dominated by a term showing $T$-linear
dependence followed by a peak developed below $T_{c}$. The onset
and the position of the peak shift to higher temperature with
decreasing the magnetic field to 5.0T (open squares in Fig. 2). If
the $T$-linear component originates from extrinsic source, it
should be persistent in the superconducting state. However,
$T_1^{-1}$ measured at low temperatures obviously goes below the
$T$-linear line shown in the figure. Then, such a case is
excluded. As for the peak the possibility of vortex motion might
be pointed out. The vortex motion has been observed from
$T_1^{-1}$ of ligand sites in high-$T_{c}$ cuprates such as
YBa$_2$Cu$_4$O$_8$ [21] or HgBa$_2$CuO$_{4+\delta}$ [22] when the
$H$ is applied perpendicular to the plane.  The FL is self-trapped
between the planes for the $H$ parallel to the plane, and thus the
effect is hardly expected in the present case. In fact, the effect
disappears when the $H$ is applied parallel to the planes in HBCO
system [22]. Furthermore, $T_1^{-1}/\gamma_N^2$ ($\gamma_N$ :
nuclear gyromagnetic ratio ) at the peak is 80 or 25 times larger
than those for YBCO or HBCO, respectively. The value is too large
to explain the peak due to vortex motion. Hence, we conclude that
the peak of $T_1^{-1}$ and the $T$-linear component have the same
origin in the electric state and/or spin fluctuations. The peak
can be assigned to a superconducting coherence peak and the
$T$-linear dependence of $T_1^{-1}$ to Korringa-type behavior,
\begin{equation}
  T_1^{-1} \propto bT
\end{equation}
 where the value of $b$ is about 6.1 (sec$^{-1}$K$^{-1}$). The
observation of the clear peak implies that a finite gap exists in
the quasi-particle excitation at all wave vectors. In the meaning
that there exist no nodes in the pairing symmetry, the
superconductivity possesses a s-wavelike character.

The spin gap is also observed in $^{63}$Cu-NMR shift ($K$) at
relevant temperatures. The shift shows $H^{-2}$-linear dependence
because large quadrupole effect acts on $^{63}$Cu nuclei. The
values free from the quadrupole effect are obtained by plotting
$K$ vs. $H^{-2}$ and extrapolating to zero field [23]. The values
at high temperatures are shown in Fig. 3. The shift is given as a
sum of two components, the orbital and the spin parts ($K=K_{orb}
+ K_{spin}$) and the spin part is proportional to the spin
susceptibility. The $T$ dependence of $K$ at high temperatures
fits well the theoretical curve for a spin-ladder system [24],
\begin{equation}
 K(T) = K_0 +  \frac{K_1}{\sqrt{T}} exp(-\Delta_{spin}/T).
\end{equation}
 The gap $\Delta_{spin}$ obtained from the fit is 217K and is comparable
with that estimated from $T_1^{-1}$. The value of $K_0$ is
estimated to be 0.25\%. The main contribution of $K_0$ comes from
$K_{orb}$ comparable with that for high-$T_{c}$ cuprates ($K_0$ of
YBCO, for example, is 0.28\% [25].) The paramagnetic contribution
corresponding to the Korringa term in $T_1^{-1}$  (Eq. (2)) should
be included in $K_0$.
\begin{figure}
\includegraphics{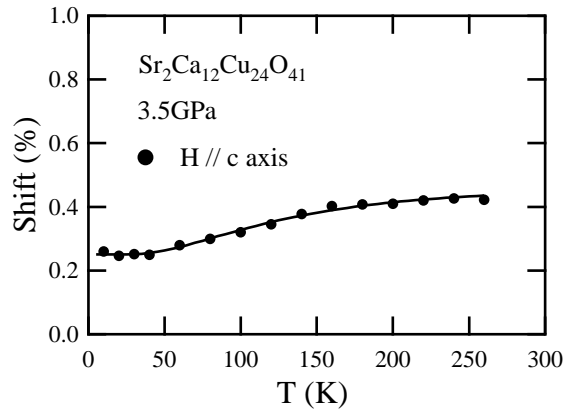}
\caption{\label{fig:epsart}  NMR shift of $^{63}$Cu nuclei for the
ladder sites at high temperatures. Solid curve represents values
calculated by Eq. (3) in the text.  }
\end{figure}
 The shift in the low
temperature region around $T_{c}$ is shown in Fig. 4.  The raw
data at several fields are also plotted in the figure since
$T_{c}$ depends on the fields. As seen from the figure, no
appreciable change is seen at $T_{c}$. The paramagnetic
contribution in $K_0$ should be considerably small so that a
change at $T_{c}$ might be difficult to detect within the present
experimental accuracy.
\begin{figure}
\includegraphics{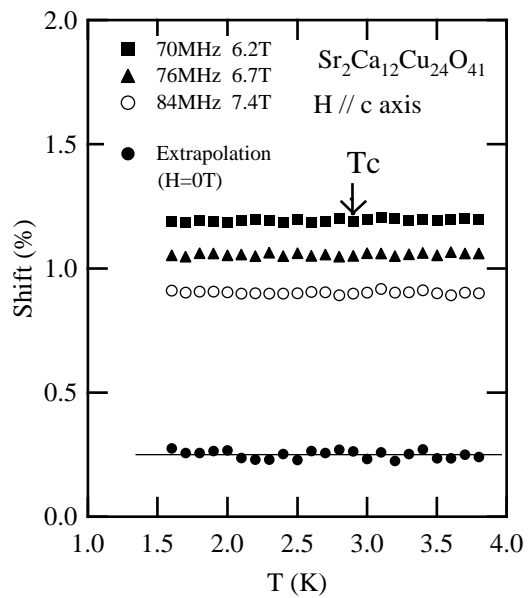}
\caption{\label{fig:epsart} NMR shift of $^{63}$Cu nuclei for the
ladder sites at low temperatures around $T_{c}$. The raw data
which show the $H^{-2}$-linear dependence due to the quadrupole
effect are also plotted in the figure. }
\end{figure}

 We have observed from $T_1^{-1}$ and $K$ unexpected features
in both normal and superconducting states. At pressure of 3.5GP we
observed two excitation modes in the normal metallic state; one
gives rise to the gapless $T$-linear component in $T_1^{-1}$(
$T_1^{-1} \propto bT$ ) which links directly with the
superconductivity, and the other the activation-type component
expressed in Eq. (1). The persistence of the spin gap at high
pressures suggests that quasi-one-dimensional spin-charge dynamics
is preserved in the normal state although pressure increases
coupling or hopping between the ladders. In the case of
conventional metals the gapless $T$-linear component arises from
paramagnetic free electrons, however, such a term is hardly
expected in the present system unless peculiarity of the ladder
structure is regarded, because majority of spins falls into the
spin-singlet state due to the existence of the large spin gap
$\Delta_{spin}$ at low temperatures.

 Since the system is metallic under high pressure, the system
 should be treated in $\it{k}$ space. However,
 microscopic snapshot in real space saves to understand the
 existence of the $T$-linear component as well as the activation-type component.
The snapshot in real space can be described as follows: The ground
state of the undoped ladders is understood as overlap of spin
dimmers on the rung [4]. Hole doping implies breaking of spin
dimmers on the rung and gives rise to holon-spinon bound state
[26]. At high temperatures singlet-triplet spin excitation in the
spin dimmers away from the bound state dominates, which causes the
activated behavior of $T_1^{-1}$ as is illustrated in Figs. 5A. At
intermediate temperatures majority of the spin dimmers falls into
the singlet ground state, but the spin in the bound states moves
rather freely and contributes to the gapless $T$-linear component
in $T_1^{-1}$(Figs. 5B).  The superconductivity would be realized
by the pairing of two bound states as illustrated in Figs. 5C. In
this viewpoint, the spin-gap formation observed from $T_1^{-1}$ at
high temperatures does not contribute to the pairing formation of
the superconductivity. The pairing force is expected to be
magnetic since the system is a strongly correlated electron
system, however, we cannot exclude the possibility that
conventional phonon coupling plays some roles in the pairing.

\begin{figure}
\includegraphics{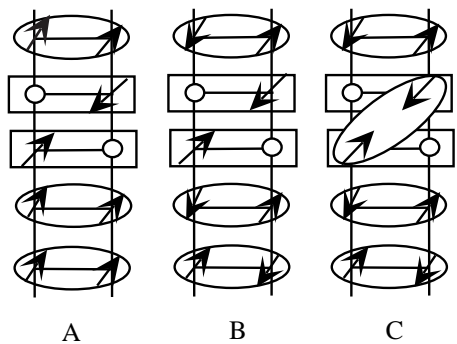}
\caption{\label{fig:wide}(A) Illustration of spin and charge
configuration at high temperatures. Ellipses show spin dimmers on
the rung. Some of them are in the triplet states at high
temperatures. Rectangle implies the holon-spinon bound states.
They move independently in the ladder.  (B) Illustration at
intermediate temperatures. A spin within the holon-spinon bound
state is free and paramagnetic. Spin dimmers in the ellipses are
in the singlet state at this temperature region. (C) Illustration
at the superconducting state. Two spins within the bound state
form the pairing.}
\end{figure}

Finally, it should be noted that the normal state in the present
system is free from large antiferromagnetic fluctuation unlike
high-$T_{c}$ cuprates. $T_1^{-1}$ in typical high-$T_{c}$ cuprates
such as YBCO systems is expressed as $a+bT$ above $T_{c}$ where
$T$-independent term $a$ represents the antiferromagnetic
fluctuation ($a \sim3$x10$^3$ (sec$^{-1}$) and $b \sim 6$
(sec$^{-1}$K$^{-1}$) for YBCO) [27]. The antiferromagnetic
fluctuation is extremely suppressed due to the existence of the
spin gap in the present system, which might explain why high
$T_{c}$ is not realized in this system although the structure is
quite similiar to high-$T_{c}$ cuprates.

In conclusion, we have succeeded in obtaining microscopic
information of the superconducting state in
Sr$_2$Ca$_{12}$Cu$_{24}$O$_{41}$  by applying pressure up to
3.5GPa. In this material, we observed two excitation modes in the
normal state. One gives rises to the activation-type component in
$T_1^{-1}$, the other $T$-linear component linking directly with
the superconductivity. The superconductivity possesses a
s-wavelike character in the meaning that a finite gap exists in
the quasi-particle excitation.

Authors wish to thank Profs. H. Fukuyama, K. Yamada, M. Imada and
M. Takigawa and Drs. S. Fujimoto, H. Kontani, and K. Kojima for
fruitful discussion. The present work was partially supported by
Grant-in-Aid for the Ministry of Education, Science and Culture,
Japan, and by grants of Ogasawara Foundation for the Promotion of
Science and Engineering, Japan and Simadzu Science Foundation,
Japan.

\end{document}